\documentclass[final,3p,times,twocolumn]{elsarticle}

\usepackage{amssymb}
\usepackage{amsmath}
\usepackage{xcolor}
\usepackage{hyperref}

\usepackage[capitalise]{cleveref}
\usepackage{siunitx}
\usepackage{tikz}
\usepackage{booktabs}

\begin{document}

\begin{frontmatter}



\title{Self-Buckling of Pressurized Cylindrical Tubes}
\author[label1]{Morten Opstrup Andersen}

\author[label1]{Nikolaj T\o{}nner Osvald Olsen}

\author[label3]{Diksha Bhola}

\author[label2]{Aleca Borsuk}

\author[label4]{Craig Brodersen}

\author[label3]{Anja Geitmann}

\author[label1]{Matteo Pezzulla\corref{cor1}}
\ead{matt@mpe.au.dk}
\cortext[cor1]{Corresponding author}

\affiliation[label1]{organization={Department of Mechanical and Production Engineering, Aarhus University},
            addressline={Katrinebjergvej 89F}, 
            city={Aarhus N},
            postcode={8200}, 
            country={Denmark}}

\affiliation[label2]{organization={New York Botanical Garden},
             addressline={2900 Southern Boulevard},
             city={Bronx},
             postcode={10458},
             state={New York},
             country={US}}

\affiliation[label3]{organization={Department of Plant Science, Faculty of Agricultural and Environmental Sciences, McGill University},
             city={Montreal},
             country={Canada}}

\affiliation[label4]{organization={School of the Environment, Yale University},
             city={New Haven},
             state={Connecticut},
             country={US}}


\begin{abstract}
We investigate the buckling of hollow cylindrical tubes subject to their own weight and internal pressure, inspired by the columnar cells of the palisade mesophyll in dicotyledon leaves which resemble pressurized cylindrical tubes. When the internal pressure in the cylinder is equal to the outside pressure, the problem is usually termed self-buckling, which has been studied extensively for solid rods, hollow cylinders, and thin cylindrical shells. Specifically, we perform FEM simulations and desktop-scale experiments to determine the instability thresholds for different geometrical parameters. We first test our models against self-buckling results without pressure for solid rods and hollow cylindrical tubes, and then proceed to determine the critical buckling pressure for a set of material and geometrical parameters. We find that positive internal pressures can stiffen cylinders that are unstable under their own weight, leading to an effective Young's modulus that we show scales linearly with the applied pressure. On the contrary, cylinders that are stable under self-weight, buckle under a negative pressure, resembling classical results on pressure-induced ring buckling. Our findings offer new insights on the interplay between gravity and pressure for the mechanical instability of hollow cylindrical tubes, which we hope will be useful for the study of both engineering and biological structures under similar loads.
\end{abstract}




\end{frontmatter}



\section{Introduction}\label{sec1}

Euler buckling is a prototypical example of elastic instability, occurring when a slender rod, for example clamped on one end, is subject to an axial compressive force on the other end. If the force exceeds a specific threshold, the rod will eventually bend \cite{euler1759}. When a rod, or a column, is in a vertical position, gravitational loads can also induce a buckling instability, known as self-buckling \cite{Greenhill}. While self-weight is usually neglected in structural or mechanical engineering, it becomes relevant for soft materials, where self-buckling can actually be used as an indirect measurement method of the Young's modulus \cite{garg2024passive}, and biology, where it governs plant allometry at multiple scales, from individual cells to entire trees \cite{niklas1994,Koch2004,Dargahi2019}.

The study of instabilities in columns dates back to Leonhard Euler with his work on column buckling \cite{euler1759,euler1744}, but
it was not until 1881 that the problem was solved by Greenhill \cite{Greenhill}. Tall columns have since been investigated with different aims and purposes, such as the study of the optimal shape to maximize the height of a solid column \cite{KellerNiordson}. More recently, Kanahama and Sato developed a method to obtain the solution for self-buckling of hollow cylinders, based on the same theory by Greenhill \cite{Kanahama}. 

Studies on self-buckling have also been focused on thin cylindrical shells, with examples dating back to 1962 \cite{Weingarten}, although more recent numerical and experimental studies can also be found \cite{Calladine1,Calladine2}. As the problem of axially compressed cylindrical shells have been studied extensively and solved analytically \cite{Flugge,Donnell}, a simplified solution for the self-buckling of cylindrical shells was based on the critical stress of the axially compressed cylinder \cite{Calladine1,Calladine2}. 

While it is important to mention studies on axially compressed cylindrical shells under pressure \cite{Hutchinson, Rotter}, and thick-walled cylinder under pressure, especially in connection with an imposed axial stretch or compression \cite{WANG1972,Hill1976,Haughton1979,Haughton1979_2}, a thorough investigation of the interplay between gravity and pressure in the buckling of hollow cylinders is lacking. Our study is motivated by the turgid palisade mesophyll cells in bifacial plant leaves, which resemble cylindrical hollow tubes. The tubular cell envelope consists of polysaccharidic cell wall material, which is under tension resulting from the hydrostatic pressure (turgor) that is controlled through osmotic processes \cite{Borsuk2024, zhang2024biomechanics}, and by other studies on the effect of turgor pressure on the deformations at a cellular level \cite{Sanati2013,Bidhendi2018}. The study is based on a combination of FEM simulations and desktop-scale experiments. First, we develop an FEM model and validate it against results on the self-buckling of rods and hollow cylinders. Then, we proceed to compare the self-buckling of hollow cylinders with the self-buckling of thin cylindrical shells, highlighting the main differences in the buckling behavior. Finally, we augment our FEM model to include internal pressure and perform desktop-scale experiments on pressurized elastomeric hollow cylinders. 

\section{Methods}

\begin{figure*}[htb]
\centering
\includegraphics[width=\linewidth]{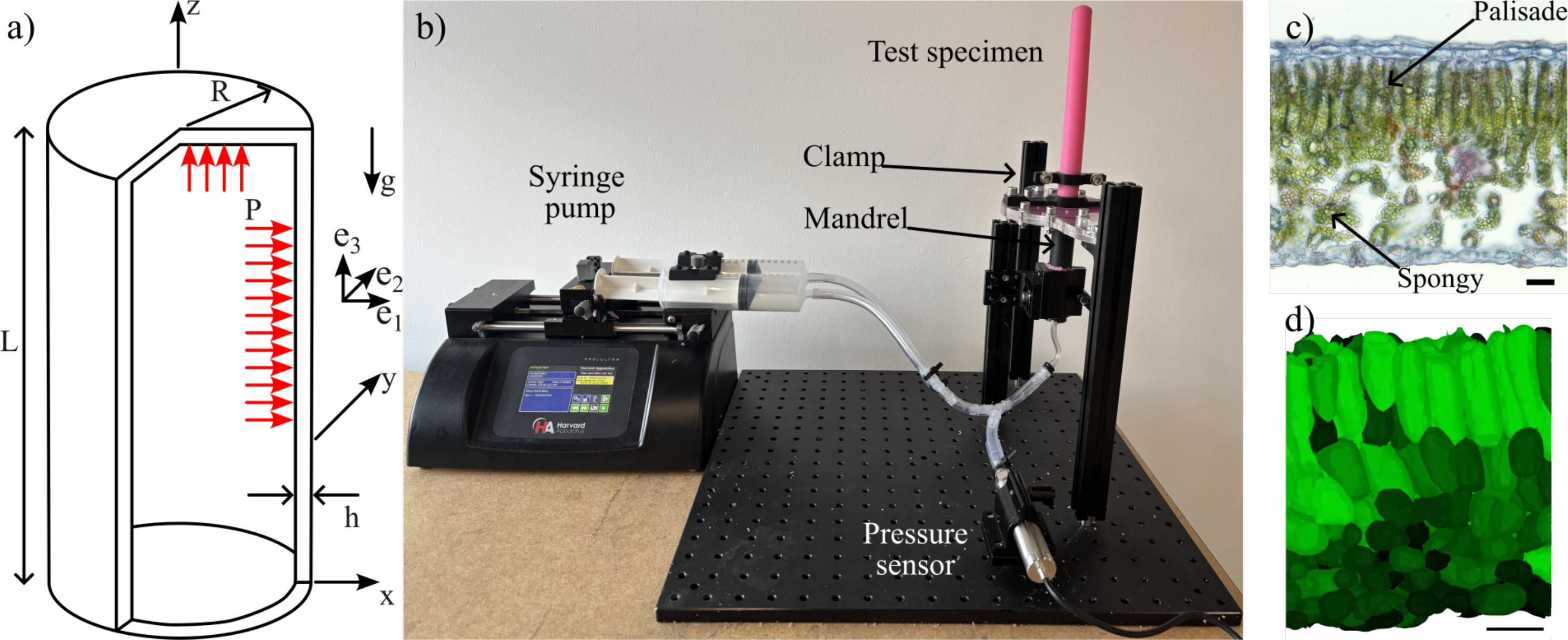}
\caption{(a) The geometry of the cylinder subjected to internal pressure $P$. (b) Experimental setup. The test specimen is clamped on an internal mandrel, which allows for volume control via a syringe pump. (c) Cross-section image of a viburnum leaf showing the internal photosynthetic tissue comprising palisade mesophyll and spongy mesophyll, with the cells pressurized through osmotic water exchange. Scale bar 20 µm. (d) 3D reconstruction of mesophyll tissue in Arabidopsis thaliana mutant \textit{cgr2-3} showing palisade cells in light green. Scale bar 50 µm.}
\label{fig:Schematics}
\end{figure*}

We will now present the experimental and numerical methods we have used to uncover the relationships between gravity, internal pressure, geometrical and material parameters, for the self-buckling of pressurized cylinders, clamped at their base. The problem we are studying is illustrated in \cref{fig:Schematics}, where a cylinder with length $L$, external radius $R$, thickness $h$, and internal pressure $P$ is subject to its own weight. If we denote by $\rho$ the volumetric density, then the specific weight can be defined as $-\rho g\mathbf{e}_3$, where $g$ is the gravitational acceleration and $\mathbf{e}_3$ is the vertical unit vector of our Cartesian basis. We also denote the Young's modulus and the Poisson ratio of the material as $E$ and $\nu$, respectively. The internal radius can be calculated as $R_\textup{i}=R-h$, and we also define the ratio between the internal and external radii as $\alpha=R_\textup{i}/R$. In our analysis, we varied $\alpha\in[0,0.95]$, $R/L\in[0.01,0.2]$, and $P/E\in[-0.08,0.09]$. Notice how we do not refer to the cylinder as cylindrical shells, as they are not necessarily thin (in most cases $\alpha$ is bounded by $0.8$ in our analysis, corresponding to $h/R=2/9$).

\subsection{Experimental methods}\label{sec:ExpMethods}

Desktop-scale experiments were performed on polymeric cylinders: both hollow and solid cylinders were manufactured via a mold-and-cast technique using VPS-8 (vinylpolysiloxane, Zhermack) on 3D printed (Bambulab P1S) cylindrical molds using basic PLA filament. The top layer, needed to seal hollow cylinders, was made by submerging the tip of the cylinder into a layer of VPS of the same thickness as the walls. We characterize VPS-8 via its Young's modulus $E=0.22\pm0.03$ MPa, a Poisson's ratio $\nu=0.49$, and a volumetric density $\rho=1070$ $\si{\kilo\gram\per\meter\cubed}$ \cite{garg2024passive}. \cref{fig:Schematics} (b) shows the experimental setup, where the cylinders are clamped on an internal mandrel, which fixes the base while introducing a method for varying their length. The mandrel has a channel that allows air to be added or removed from the internal void of the cylinders. The channel is connected to a $100$ bar pressure sensor (Omega, PXM409-100BGUSBH) and a syringe pump (Harvard Apparatus PHD Ultra 70-3007) with two 150 mL syringes, which allows for volume-controlled experiments \cite{Yan2020}. 
The deformation of the cylinder was recorded via a Basler camera (acA4096-40uc with an Edmund Optics 25mm/F1.4 lens). These recordings were used to determine the buckling events, which we will describe later.

\subsection{Computational framework}\label{ssec:compfw}



To investigate the problem numerically, we resort to the finite element method in the commercial software COMSOL Multiphysics v. 6.2, with a geometrically non-linear formulation. In using the finite element method to find the critical points for non-linear buckling problems, popular choices include the arc-length method \cite{RIKS} and dynamic relaxation \cite{Lavrencic2018}. This work uses the dynamic approach, by employing an implicit time-stepping scheme in COMSOL Multiphysics. Our simulations were performed with a second-order backward differentiation formulation (BDF2), without any additional damping except that inherent to implicit time-integration methods.

We use a structured mesh with quadratic Lagrangian shape functions, while convergence was studied by increasing the number of elements around the circumference of the cylinders.
Although our mesh is structured, we insert an element vertex placed at $(x,y)=(R,0)$ and $4.5$ elements above $z=0$, to introduce a controlled asymmetry, leading to consistent buckling paths under our implicit implementation. To be able to study the elastic instability when gravity is coupled with pressure, we use an imperfection in the form of a slightly tilted gravitational field, such that its component in the negative y-direction is $g/1000$.

We use two types of material models based on the Saint Venant-Kirchhoff (SVK) energy \cite{Ogden1985}, valid for small strains, and the nearly-incompressible neo-Hookean energy \cite{Ogden1985,NeoHook}. Although strains in our problem are small and the Saint Venant-Kirchhoff model suffices, we also consider the neo-Hookean model to verify that strains are indeed small and to assess whether there are any significant differences between the two. The Saint Venant-Kirchhoff model is implemented within the Solid Mechanics interface, where a linear material model is selected. On the other hand, the neo-Hookean model is implemented via the weak-form interface, following \cite{NeoHook,Ogden1985}. In the calculations, values of unity are chosen for the length $L$ and Young's modulus $E$, while for the nearly-incompressible neo-Hookean model, we use the shear modulus $\mu=E/3$ and bulk modulus $\kappa=10000\mu$. The pressure forces are imposed on the internal boundary of the cylinder (both lateral and top surfaces) as live surface loads, and all cylinders are clamped at the bottom surface where the displacement field is set to zero $\mathbf{u}=\mathbf{0}$. 

To study the instability in pressurized cylinders, we add static steps to our numerical procedure. Specifically, in the negative pressure regime, one static step is added to apply the gravitational load before the pressure is decreased over time. In the positive pressure regime, two static steps are added: first, a positive pressure is applied as a static step before the second static step, where the gravitational load is introduced. Finally, the critical combinations of $P/E$, $R/L$, and $\alpha$ are then found by decreasing the pressure until buckling occurs.

\subsection*{Non-dimensionalization}
The numerical models are implemented in a dimensionless form to identify the key parameters and reduce the computational cost during the parametric analyses. We non-dimensionalize lengths and coordinates by $L$, stresses, $\sigma$, and pressure, $P$, by $E$, by introducing the following relationships 
\begin{equation}
    x=L\tilde{x}\,,\quad \sigma=E\tilde{\sigma}\,,
\end{equation}
where $\tilde{\cdot}$ denote dimensionless variables. The dimensionless pressure will be written as $P/E$ in the remainder of the paper. Plugging these into the stationary equations of motion
\begin{equation}\label{eq:NDbalance}
    \tilde{\sigma}_{ij,j}+\frac{L}{E}f_i=0\,,
\end{equation}
where $\sigma_{ij}$ are the Cartesian components of the Cauchy stress tensor, $(\cdot),_j$ denote differentiation with respect to the j-th coordinate, and $f_i$ are the Cartesian components of the body forces that are all zero except for $i=3$. We then redefine $(L/{E})f_3$ as 

\begin{equation}\label{eq:beta}
    \beta=\frac{L}{E}\rho g \,.
\end{equation}
We notice that, in the numerical implementation, we introduce the imperfection by setting the component along $\mathbf{e}_2$ to be nonzero, specifically $f_2=-0.001\rho g$.

The dimensionless parameter $\beta$ is consistent with that used in \cite{Calladine2}, and is indicative of the body weight relative to the stiffness of the structure. As $\beta$ increases, for a given $\alpha$ and $R/L$, the structure will become more susceptible to buckling under its own weight. Once the critical value of $\beta$ is determined, one can obtain the critical length (the maximum height of the cylinder) that triggers self-buckling for a given set of material and geometrical parameters. When studying the problem of self-buckling with internal pressure, the critical value of $\beta$ will change as the internal pressure varies. 

\subsection*{Determination of the buckling threshold}
The buckling threshold is determined by analyzing the average vertical displacement of the top cross section $w_\textup{t}=\mathbf{u}\cdot\mathbf{e}_3$ as a function of the imposed load, that is $\beta$. 
For the case without pressure, the vertical displacement will be linear with $\beta$, before buckling, such that the ratio $w_\textup{t}/\beta$ is constant. When approaching buckling, the vertical displacement will start to grow faster than $\beta$, leading to a deviation of the ratio from this constant. We then define the buckling threshold $\beta_\textup{cr}$ as the value of $\beta$ corresponding to a relative increase of $2\%$ for $w_\textup{t}/\beta$ from the constant value in the pre-buckling regime.

For cylinders with positive internal pressure, two distinct instability mechanisms are observed numerically as the pressure is reduced. First, a bending instability of the initially straight cylinder emerges smoothly, analogous to classical buckling in the absence of pressure. In contrast to the unpressurized case, however, this bent configuration remains stable and corresponds to a static equilibrium state: if the pressure is held constant, the cylinder remains stationary. Upon further pressure reduction, a subcritical instability occurs, leading to the formation of a localized kink near the base and the eventual collapse of the cylinder. The collapse produces a more pronounced change in the response, which is reliably detected through the time derivative of the enclosed volume, whose slope exhibits a sharp variation at the point of collapse. This criterion, based on the time derivative of the volume, is also adopted to identify buckling under negative internal pressures.

\section{Results}
\subsection*{Self-buckling}

The problem of self-buckling for elastic rods has been solved by Greenhill \cite{Greenhill}, and it has been more recently extended to hollow cylinders by Kanahama et al. \cite{Kanahama}. We note here that a solid cylinder can be made more resistant to self-buckling, \textit{i.e.} its critical $\beta$ can be increased, by making it hollow while keeping the external radius constant. This difference can be explained by looking at the ratio between the moment of inertia $I$ and the cross-sectional area $A$, which dictates the maximum compressive stress a column can sustain before buckling \cite{Greenhill,euler1744}. As the moment of inertia of the cross-section of a hollow cylinder scales as $I\propto (1-\alpha^4)R^4$ while its area scales as $A\propto (1-\alpha^2)R^2$, the ratio $I/A$ scales as $I/A\propto (1+\alpha^2)R^2$, highlighting a monotonic increase with $\alpha$, as exactly derived by Kanahama et al. \cite{Kanahama}. More precisely, the critical length of hollow cylinders is determined as
\begin{equation}\label{eq:LcritCubed}
    L_\textup{cr}^3=\frac{2ER^2}{\rho g}(1+\alpha^2)\,,
\end{equation}
which can also be combined with the non-dimensionalization to obtain $\beta_\textup{H}/\beta_\textup{S}=(1+\alpha^2)^{1/3}$, where the subscripts $H$ and $S$ refer to hollow and solid cylinders, respectively.

It is important to point out that while Kanahama's results provide an equation for the critical length as a function of $\alpha$, they are not expected to hold when $\alpha\rightarrow1$, as hollow rods transition to cylindrical shells, since the model is based on Euler-Bernoulli beam theory \cite{Kanahama}. 
The beam theory is not appropriate to model thin shells, as shells are characterized by a coupling between stretching and bending energies of the mid-surface, which effectively restricts changes in the Gaussian curvature, forcing the deformations to localize \cite{Calladine1983ToSS,Miura2020,Qiu_2022}. 

As a validation step for our numerical model, we carried out simulations of the self-buckling problem. In this case, cylinders are modeled without a top lid, since a pressure differential does not need to be maintained, and neither the Greenhill nor Kanahama theories include it. In our dimensionless numerical model, for each value of $R/L$ and $\alpha$, we increase the parameter $\beta$ linearly with time until we observe buckling, as detailed previously. For this study, we considered both neo-Hookean and SVK strain energies, with a Poisson ratio equal to $\nu=0$ and $\nu=0.49$ for the latter. 

\begin{figure}[t]
    \centering
    \includegraphics[width=\linewidth]{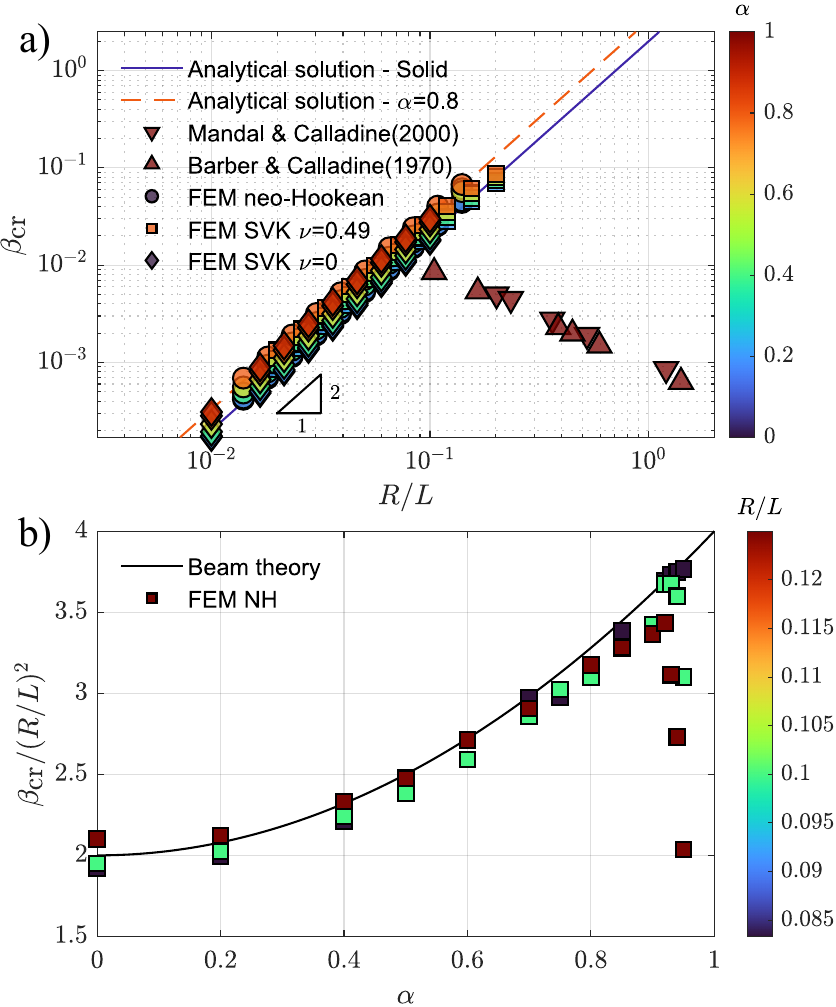}

    \caption{Self-buckling of solid and hollow cylinders. The critical value of $\beta_\textup{cr}$ is shown as a function of $R/L$, while the colorbar indicates different values of $\alpha$. Solid black and dashed blue lines represent the analytical solution for $\alpha=0$ and $\alpha=0.8$, respectively. Solid symbols represent our numerical results for the SVK energy with $\nu=0$ (diamonds) and $\nu=0.49$ (squares), and for the neo-Hookean energy (circles). Triangles are experimental results on cylindrical shells from \cite{Calladine1,Calladine2}, to highlight the different behavior between hollow cylinders and cylindrical shells.}
    \label{fig:self_buckling}
\end{figure}

\cref{fig:self_buckling} shows the critical buckling threshold $\beta_\textup{cr}$ as a function of $R/L$, as obtained via our FEM simulations for SVK energy with $\nu=0$ (diamonds) and $\nu=0.49$ (squares), and neo-Hookean energy (circles). The theoretical prediction by Kanahama is recast in terms of $\beta$ as
\begin{equation}\label{eq:betaC}
    \beta_\textup{cr}=2\left(1+\alpha^2\right)\left(\frac{R}{L}\right)^2\,,
\end{equation}
and plotted as a solid line for $\alpha=0$ (blue) and dashed line for $\alpha=0.8$ (orange). We find good agreement between our simulations and the analytical solutions for both material models, thereby validating our numerical approaches and confirming that strains are indeed small. Moreover, \cref{fig:self_buckling} (a) shows experimental results on the self-buckling for cylindrical shells with $\alpha\in[0.987, 0.998]$ from \cite{Calladine1,Calladine2}. 

We notice how experimental results on thin shells do not follow the quadratic scaling with $R/L$ as the buckling behavior is completely different: while rods or hollow cylinders buckle by realizing a bending mode \cite{Greenhill,euler1744,Kanahama}, thin shells buckle either by crumpling at the base or by a localized bending of the mid-surface near the top of the cylinder, depending on their slenderness. This means that the buckling strength for a fixed $R/L$ is non-monotonic in $\alpha$: while a hollow cylinder can sustain a higher $\beta$ than a solid one, its buckling strength will drop significantly as $\alpha\rightarrow1$, a behavior that cannot be predicted via a beam theory.
This non-monotonic behavior of $\beta_\textup{cr}$ versus $\alpha$ is highlighted in \cref{fig:self_buckling} (b), where a study with $R/L=\{1/8, 1/10,1/12\}$ is reported. Specifically, we plot $\beta_\textup{cr}/(R/L)^2$ as a function of $\alpha$ and observe a collapse of the data for different $R/L$ and its non-monotonic pattern. The solid line is derived from \cref{eq:betaC}, once divided by $(R/L)^2$, to give $\beta_\textup{cr}/(R/L)^2=2(1+\alpha^2)$. 

We can also observe how shells with a higher $R/L$ display a significantly larger drop in $\beta_\textup{cr}$, highlighting how the critical $\beta$ for thin cylindrical shells also depend on $R/L$. This is in line with suggestions by John W. Hutchinson for the paper by Barber and Calladine to perform experiments on cylinders with a different diameter, a task that Mandal and Calladine later carried out \cite{Calladine3}.

\subsection{Self-buckling under pressure}

To study the self-buckling of hollow cylinders subject to internal pressure, we modified the geometry by adding a top lid, as thick as the walls of the cylinders. To evaluate the impact of the thickness of the top lid on our problem, we ran FEM simulations on the self-buckling without pressure for different thicknesses of the top lid. As our results on the self-buckling in \cref{fig:self_buckling} show that there are no significant differences between SVK and neo-Hookean FEM results, we only used the SVK model. We found that the top lid reduces $\beta_\textup{cr}$ by $2.9\%$ for hollow cylinders with $R/L=1/20$ and by $5.9\%$ for $R/L=1/10$. These values are averaged from data with $\alpha\in[1/2, 3/4]$. We note that the impact of adding a top becomes more significant for higher values of $\alpha$. 
\begin{figure}[tb]
\includegraphics[width=\linewidth]{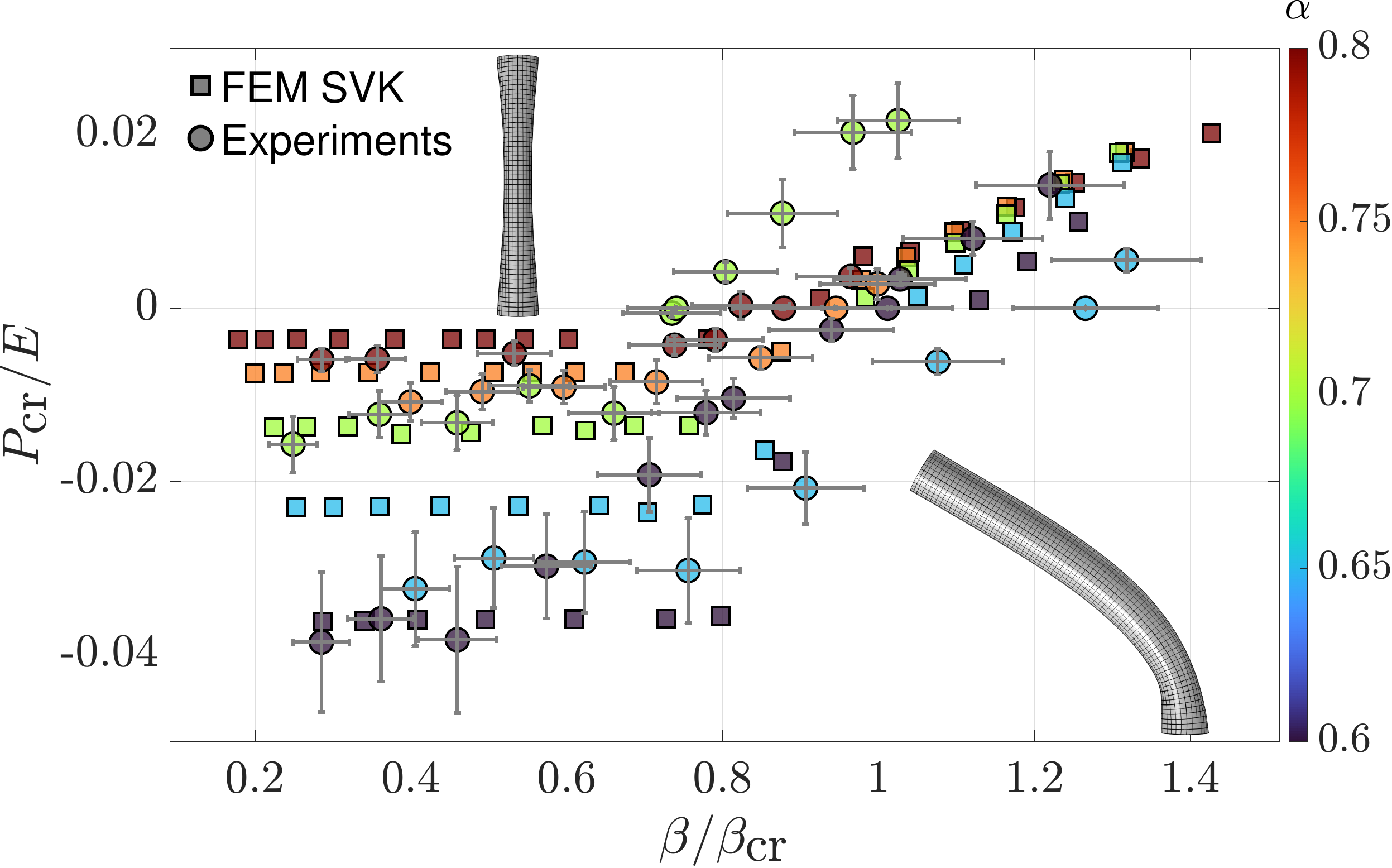}
\caption{Critical dimensionless pressure, $P_\textup{cr}/E$, as a function of $\beta/\beta_\textup{cr}$, for different values of $\alpha$ as shown in the colorbar. Numerical results are shown as solid squares, while experiments are shown as circles with corresponding error bars. The two insets show the characteristic buckling modes for $\beta/\beta_\textup{cr}>1$ (bending), and for $\beta/\beta_\textup{cr}<1$ (localized buckling).}
\label{fig:REvsRL}
\end{figure}

\cref{fig:REvsRL} shows the critical dimensionless pressure, $P_\textup{cr}/E$, as a function of $\beta/\beta_\textup{cr}$, for $\alpha\in[0.6,0.8]$. Here, $\beta_\textup{cr}$ is the critical self-buckling threshold without pressure ($P=0$). Indeed, we observe that $P_\textup{cr}/E=0$ for $\beta/\beta_\textup{cr}=1$, which corresponds to pure self-buckling. For $\beta/\beta_\textup{cr}>1$, cylinders are unstable under their own weight and a positive pressure can stabilize them, in line with previous observations on thick-walled cylinders with internal pressure under compressive loads \cite{Haughton1979,Haughton1979_2}. Indeed, as noted in the Methods, two instabilities could be identified numerically in this regime when lowering the internal pressure from a high enough value that stabilizes the cylinders: a first super-critical buckling behavior, where the structure starts bending from the initial straight configuration, and a second collapse of the cylinder with a kink forming near the base. While the latter can be identified both numerically and experimentally, the former cannot be characterized in our experiments, due to initial misalignment and imperfections in the system. We will therefore proceed to characterize the catastrophic buckling instability. Our numerical results show a linear relationship between $P_\textup{cr}/E$ and $\beta/\beta_\textup{cr}$, and they qualitatively agree with our experimental results, although a quantitative agreement in this regime across all values of $\alpha$ is lacking. This might be due to various imperfections, such as a varying wall thickness, especially for the top lid, misalignment of the test specimen, and a not perfect clamp. However, we can observe how the experimental and numerical results in \cref{fig:REvsRL} show roughly the same slope.

\begin{figure}[tb]
    \centering
    \includegraphics[width=\linewidth]{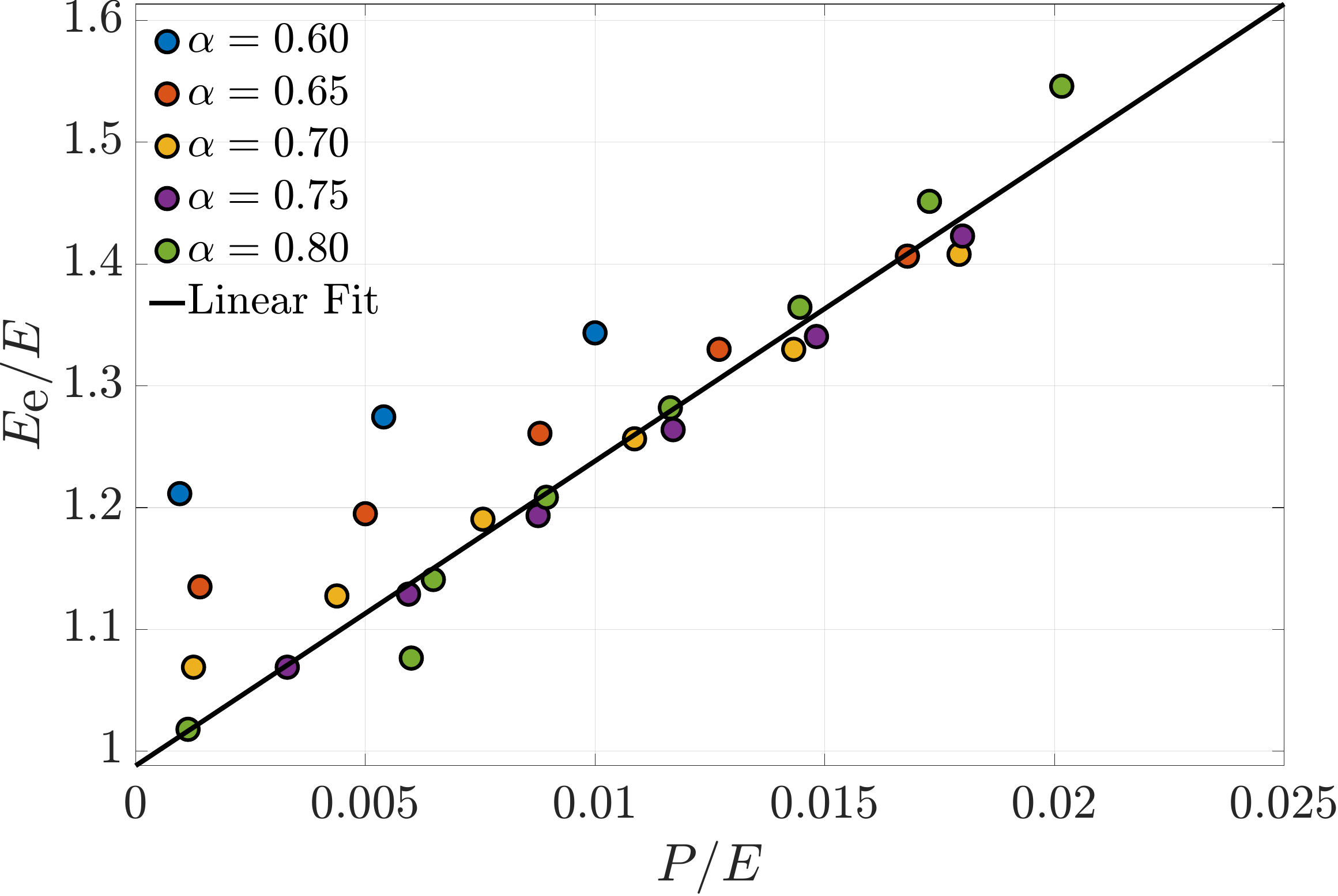}
    \caption{Effective Young's modulus normalized by the Young's modulus of the base material, $E_\textup{e}/E$ as a function of $P/E$ for cylinders subject to positive pressure, for $\alpha\in[0.6,0.8]$. Regression is performed for $\alpha\in[0.7,0.8]$, which yields the solid black line.}
    \label{fig:EvsPE}
\end{figure}

Given the observed linear relationship between $P_\textup{cr}/E$ and $\beta/\beta_\textup{cr}$, we set to obtain the effective Young's modulus of pressurized cylinders in the form $E_\textup{e}=E+aP$, with $a\in\mathbb{R}$. This linear relationship is also inspired by previous studies on the relation between turgor pressure and tissue rigidity in plants \cite{Nilsson1958}. Although thick-walled cylinders under internal pressure and tension have been previously investigated \cite{Hill1976,Haughton1979, Haughton1979_2}, none of these studies account for the structure's self-weight. To calculate the effective modulus $E_\textup{e}$ at buckling, we make use of our FEM simulations. Specifically, we look for the minimum value of the Young's modulus, which we term $E_\textup{e}$, such that the hollow cylinder is stable, without internal pressure. This higher Young's modulus allows the cylinders that would otherwise buckle without internal pressure to remain stable. We compute this for each combination of $\alpha$ and $\beta/\beta_\textup{cr}$, and then combine the results with $P_\textup{cr}/E$ versus $\beta/\beta_\textup{cr}$ to find a relationship of the effective Young's modulus as a function on the internal pressure. \cref{fig:EvsPE} shows the dimensionless effective Young's modulus, $E_\textup{e}/E$ as a function of $P/E$, for different values of $\alpha$. We can observe a common linear trend between $E_\textup{e}/E$ and $P/E$ for $\alpha>0.7$, while lower values of $\alpha$ deviate from it, even suggesting that $E_\textup{e}/E\ne1$ for $P=0$. This discrepancy might be attributed to different buckling modes between the original pressurized problem and the equivalent unpressurized one. The linear fit, plotted in \cref{fig:EvsPE}, is given by $E_\textup{e}/E=25P/E+1$. 

\begin{figure}[tb]
    \centering
    \includegraphics[width=\linewidth]{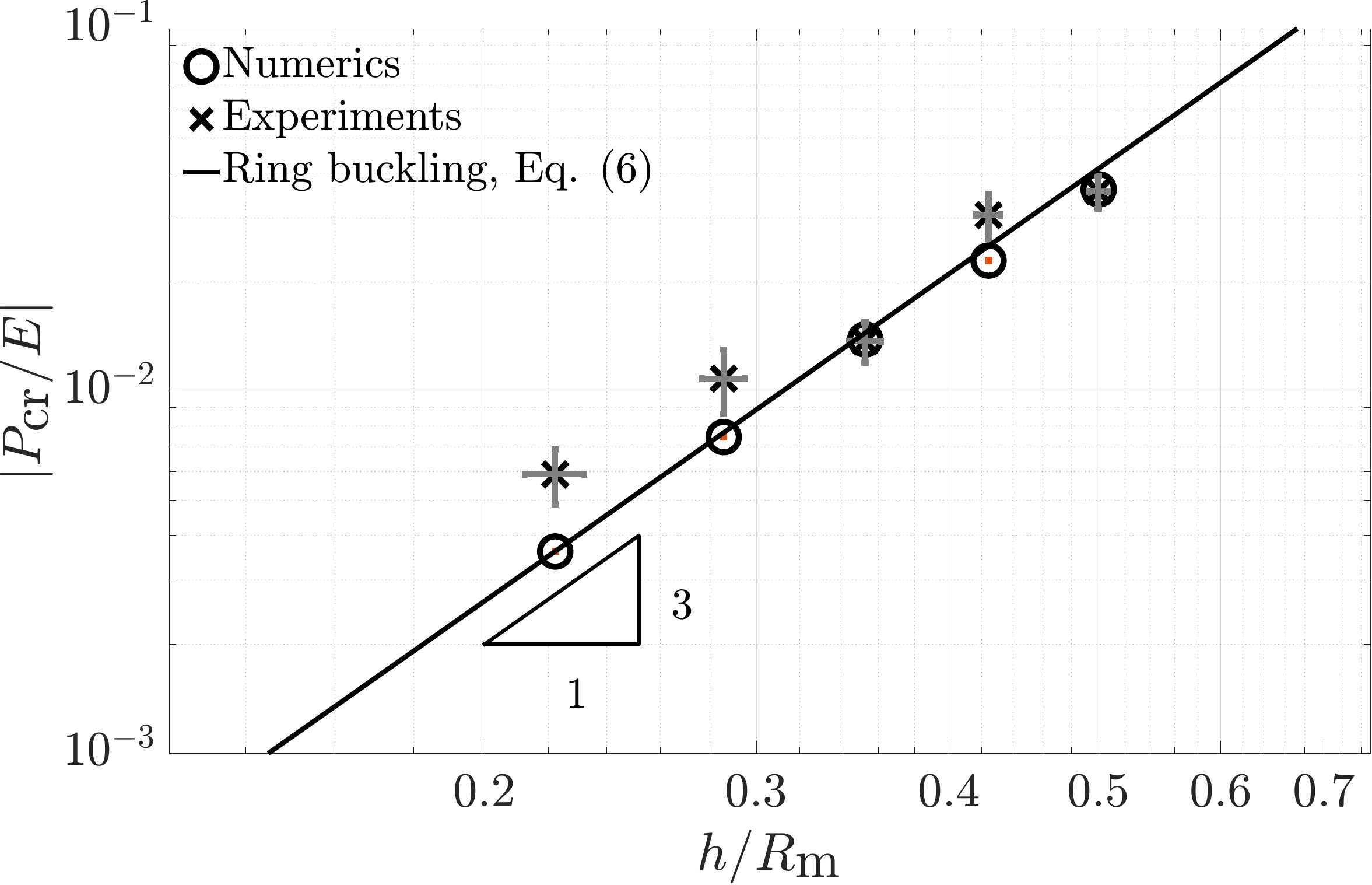}
    \caption{Absolute value of the critical, dimensionless, pressure, $P_\textup{cr}/E$, in the plateau regime versus $h/R_\textup{m}$. Experiments are represented as crosses, while numerical results are represented as circles. Error bars denote the standard deviation of the mean. The solid black line represents the theoretical prediction of the critical pressure for ring buckling.}
    \label{fig:Ring}
\end{figure}

We now turn our attention to the plateau region in \cref{fig:REvsRL}, where self-weight has a negligible influence on the buckling behavior. Our FEM simulations and experiments are in good agreement with each other in this region, although we observe some fluctuations around the average values for each $\alpha$. Therefore, we averaged the values for $P_\textup{cr}/E$ for both experiments and simulations over the plateau region with $\beta/\beta_\textup{cr}<0.8$, and obtained the results presented in \cref{fig:Ring} in terms of the absolute value of $P_\textup{cr}/E$ versus $h/R_\textup{m}$, where $R_\textup{m}=(R_\textup{i}+R)/2$ is the radius of the mid-surface. Error bars for the numerical simulations represent the standard error of the mean and are smaller than the symbol size. Experimental uncertainties on $R$ are propagated to generate error bars for $h/R_\textup{m}$, while the error bars for $P_\textup{cr}/E$ are obtained by propagating the uncertainties of the individual measurements through the arithmetic mean. This plot highlights a cubic scaling of the critical (dimensionless) pressure with $h/R_\textup{m}$, in agreement with the buckling behavior of thin, long cylindrical shells under negative internal pressure and for rings under external pressure, for which a linear stability analysis leads to \cite{TimoshenkoGere1961,Singer1998}
\begin{equation}\label{eq:PEring}
    P_\textup{cr}=\frac{E}{4(1-\nu^2)} \left(\frac{h}{R_\textup{m}}\right)^3\,.
\end{equation}  
Indeed, as also the deformed shape in the top-right of \cref{fig:REvsRL} shows, the cylinders in this regime buckle via a localized deformation at around half of their height, suggesting that self-weight is indeed negligible in this regime. This thin-walled solution has also previously been shown to closely match the results for thick-walled cylinders of a neo-Hookean material without self-weight \cite{WANG1972}, and is described in relation to the collapse of mesophyll cells with negative turgor pressure \cite{Ding2014}.

\section{Discussion}

Our results uncover the interplay between internal pressure and self-weight in the buckling of hollow cylinders, inspired by the mesophyll cells with turgor pressure in leaves, which resemble cylindrical pressurized tubes. Our analyses show that an internal positive pressure can stabilize hollow cylinders, which would otherwise buckle under their own weight. Via FEM simulations, we have also derived an effective Young's modulus for pressurized hollow cylinders, to estimate how the stiffness of the structure can be increased via internal pressure.

In biological contexts, such as mesophyll cells, this suggests that turgor pressure may therefore serve not only to maintain shape but also to modulate effective stiffness and resistance to gravity-driven deformation \cite{Nilsson1958}. Although negative turgor pressures might seem unlikely in hydrated and living cells, they can occur as a result of drought or exposure to high osmotic medium. Experimental and theoretical studies suggest they may occur particularly when cell walls are sufficiently stiff, which would limit buckling \cite{Ding2014}. Similarly, it has been shown that the negative pressures arising in xylem sap are compensated by development of cylindrical xylem conduits with appropriate thickness-to-radius ratios that prevent implosion \cite{pittermann2006mechanical}. 

Our results extend this perspective by showing how the coupling between pressure and self-weight further constrains the range of stable geometries and may help explain why specific cellular aspect ratios are observed in Nature. We calculated the $R/L$ ratio for palisade mesophyll cells from $30$ different species reported in the literature, with a range from $0.059$ to $1.198$, and a mean value of $0.273$, all within the range of values considered here (see Supplementary Material). 


Finally, our work raises several open questions. One concerns the extent to which the effective modulus $E_\textup{e}$ inferred from buckling can be used inversely to estimate internal pressures in biological cells or soft structures when geometric parameters and wall thickness are known \cite{VellaDominic2012}. Another is whether our analysis can be extended to the study of the entire mesophyll, where several cylindrical cells are in contact with one another. Addressing these questions and improving our experimental setup would further bridge the gap between continuum mechanics, plant biomechanics, and the design of pressurized soft systems.

\section*{CRediT authorship contribution statement}
\textbf{Morten Opstrup Andersen:} Writing - review and editing, Writing - original draft, Visualization, Validation, Methodology, Formal analysis, Conceptualization.

\textbf{Nikolaj T\o{}nner Osvald Olsen:}
Writing - review and editing, Validation, Formal analysis, Investigation.

\textbf{Diksha Bhola:}
Writing - review and editing, Visualization 

\textbf{Aleca Borsuk:}
Writing - review and editing, Visualization 

\textbf{Craig Brodersen:}
Writing - review and editing, Visualization 

\textbf{Anja Geitmann:}
Writing - review and editing, Visualization 

\textbf{Matteo Pezzulla:}
Writing - review and editing, Methodology, Conceptualization, Supervision.

\section*{Declaration of competing interest}
The authors declare that they have no known competing financial interests or personal relationships that could have appeared to
influence the work reported in this paper

\section*{Acknowledgments}
This work was supported by a Research Grant from HFSP (Ref.-No: RGP010/2023).

\section*{Data availability}
The data and code generated during the current study are available
from the corresponding author upon request.

\bibliographystyle{elsarticle-num} 
\bibliography{bibliography}

@book{euler1744,
author = {Euler, Leonhard},
booktitle = {Methodus inviendi lineas curvas maximi minimive proprietate gaudentes, sive, Solutio problematis isoperimetrici latissimo sensu accepti},
publisher = {M.-M. Bousquet},
series = {Landmarks of Science},
title = {Methodus inviendi lineas curvas maximi minimive proprietate gaudentes, sive, Solutio problematis isoperimetrici latissimo sensu accepti [microform] },
year = {1744},
}

@article{euler1759,
  title={Sur la force des colonnes},
  author={Euler, Leonhard},
  journal={Memoires de l'Academie des Sciences de Berlin},
  pages={252-282},
  year={1759},
  volume={13}
}

@article{Greenhill,
author={Greenhill, A. G.},
title={Determination of the greatest height consistent with stability that a vertical pole or mast can be made, and the greatest height to which a tree of given proportions can grow},
journal={Proceedings of the Cambridge Philosophical Society},
year={1881},
volume={4},
pages={65-73},
doi={10.1051/jphystap:018820010033701}
}

@article{KellerNiordson,
pages = {433-446},
title = {The Tallest Column},
volume = {16},
year = {1966},
author = {Keller, Joseph and Niordson, Frithiof},
journal = {Indiana University Mathematics Journal},
number = {5},
doi = {10.1512/iumj.1967.16.16029},
}

@article{Kanahama,
title = {Summation rules in critical self-buckling states of cylinders},
journal = {Mechanics Research Communications},
volume = {123},
pages = {103905},
year = {2022},
doi = {10.1016/j.mechrescom.2022.103905},
author = {Tohya Kanahama and Motohiro Sato},
}

@article{Dargahi2019,
    author = {Dargahi, Mojtaba and Newson, Timothy and Moore, John},
    title = {Buckling behaviour of trees under self-weight loading},
    journal = {Forestry: An International Journal of Forest Research},
    volume = {92},
    number = {4},
    pages = {393-405},
    year = {2019},
    month = {05},
    doi = {10.1093/forestry/cpz027},
}

@article{Donnell,
    author = {Donnell, L. H.},
    title = {A New Theory for the Buckling of Thin Cylinders Under Axial Compression and Bending},
    journal = {Transactions of the American Society of Mechanical Engineers},
    volume = {56},
    number = {8},
    pages = {795-806},
    year = {1934},
    doi = {10.1115/1.4019867},
}

@article{Calladine1,
    author = {Calladine, C. R. and Barber, J. N.},
    title = "{Simple Experiments on Self-Weight Buckling of Open Cylindrical Shells}",
    journal = {Journal of Applied Mechanics},
    volume = {37},
    number = {4},
    pages = {1150-1151},
    year = {1970},
    month = {12},
    doi = {10.1115/1.3408677},
}

@article{Calladine2,
author = {Mandal, P and Calladine, C.R},
journal = {International journal of solids and structures},
number = {33},
pages = {4509-4525},
title = {Buckling of thin cylindrical shells under axial compression},
volume = {37},
year = {2000},
doi={10.1016/S0020-7683(99)00160-2}
}

@article{Calladine3,
author = {Calladine, Christopher R},
journal = {Advances in structural engineering},
number = {16},
pages = {2393-2403},
title = {Shell buckling, without ‘imperfections’},
volume = {21},
year = {2018},
doi={10.1177/1369433217751585}
}

@book{Calladine1983ToSS,
publisher = {Cambridge University Press},
title = {Theory of Shell Structures},
year = {1983},
author = {Calladine, C. R.},
}

@article{Flugge,
author = {Fl{\"{u}}gge, W.},
journal = {Ingenieur-Archiv},
number = {5},
pages = {463-506},
title = {Die Stabilität der Kreiszylinderschale},
volume = {3},
year = {1932},
doi={10.1007/BF02079822}
}

@article{Weingarten,
    author = {Weingarten, V. I.},
    title = "{The Buckling of Cylindrical Shells Under Longitudinally Varying Loads}",
    journal = {Journal of Applied Mechanics},
    volume = {29},
    number = {1},
    pages = {81-85},
    year = {1962},
    month = {03},
    doi = {10.1115/1.3636502},
}

@article{Hutchinson,
author = {Hutchinson, John},
title = {Axial buckling of pressurized imperfect cylindrical shells},
journal = {AIAA Journal},
volume = {3},
number = {8},
pages = {1461-1466},
year = {1965},
doi = {10.2514/3.3169},
}

@article{Rotter,
author = {Rotter, J. Michael},
title = {Elephant's foot buckling in pressurised cylindrical shells},
journal = {Stahlbau},
volume = {75},
number = {9},
pages = {742-747},
doi = {10.1002/stab.200610079},
year = {2006},
}

@article{VellaDominic2012,
author = {Vella, Dominic and Ajdari, Amin and Vaziri, Ashkan and Boudaoud, Arezki},
journal = {Journal of the Royal Society interface},
number = {68},
pages = {448-455},
title = {The indentation of pressurized elastic shells: from polymeric capsules to yeast cells},
volume = {9},
year = {2012},
doi={10.1098/rsif.2011.0352}
}

@article{Qiu_2022,
   title={Bending Instability of Rod-Shaped Bacteria},
   volume={128},
   doi={10.1103/PhysRevLett.128.058101},
   number={5},
   journal={Physical Review Letters},
   author={Qiu, Luyi and Hutchinson, John W. and Amir, Ariel},
   year={2022}, 
}

@article{RIKS,
title = {An incremental approach to the solution of snapping and buckling problems},
journal = {International Journal of Solids and Structures},
volume = {15},
number = {7},
pages = {529-551},
year = {1979},
doi = {10.1016/0020-7683(79)90081-7},
author = {E. Riks},
}

@article{Lavrencic2018,
title = {Simulation of shell buckling by implicit dynamics and numerically dissipative schemes},
journal = {Thin-Walled Structures},
volume = {132},
pages = {682-699},
year = {2018},
issn = {0263-8231},
doi = {10.1016/j.tws.2018.08.010},
author = {Marko Lavrenčič and Boštjan Brank},
}

@article{NeoHook,
author = {Adler, J. H. and Dorfmann, L. and Han, D. and MacLachlan, S. and Paetsch, C.},
journal = {IMA journal of applied mathematics},
number = {5},
pages = {889-914},
publisher = {Oxford Univ Press},
title = {Mathematical and computational models of incompressible materials subject to shear},
volume = {79},
year = {2014},
doi = {10.1093/imamat/hxu022},
}

@article{Nilsson1958,
author = {Nilsson, S. Bertil and Hertz, C. Hellmuth and Falk, Stig},
journal = {Physiologia plantarum},
number = {4},
pages = {818-837},
title = {On the Relation between Turgor Pressure and Tissue Rigidity. II: Theoretical Calculations on Model Systems},
volume = {11},
year = {1958},
doi = {10.1111/j.1399-3054.1958.tb08275.x},
}

@book{Miura2020,
author = {Miura, Koryo},
edition = {1},
isbn = {9780521432740},
publisher = {Cambridge University Press},
title = {Forms and Concepts for Lightweight Structures},
year = {2020},
}

@article{garg2024passive,
  title={Passive viscous flow selection via fluid-induced buckling},
  author={Garg, Hemanshul and Ledda, Pier Giuseppe and Pedersen, Jon Skov and Pezzulla, Matteo},
  journal={Phys. Rev. Lett.},
  volume={133},
  number={8},
  pages={084001},
  year={2024},
  publisher={APS},
  doi={10.1103/PhysRevLett.133.084001}
}

@article{WANG1972,
title = {Stability and vibrations of elastic thick-walled cylindrical and spherical shells subjected to pressure},
journal = {International Journal of Non-Linear Mechanics},
volume = {7},
number = {5},
pages = {539-555},
year = {1972},
doi = {10.1016/0020-7462(72)90043-1},
author = {A.S.D. Wang and A. Ertepinar},
}

@article{Hill1976,
author = {Hill, James M.},
journal = {Journal of Elasticity},
number = {2},
pages = {113-123},
title = {Closed form solutions for small deformations superimposed upon the simultaneous inflation and extension of a cylindrical tube},
volume = {6},
year = {1976},
doi = {10.1007/BF00041780},
}

@article{Koch2004,
author = {Koch, George W. and Sillett, Stephen C. and Jennings, Gregory M. and Davis, Stephen D.},
journal = {Nature},
number = {6985},
pages = {851-854},
title = {The limits to tree height},
volume = {428},
year = {2004},
doi = {10.1038/nature02417},
}

@book{niklas1994,
  title={Plant Allometry: The Scaling of Form and Process},
  author={Niklas, K.J.},
  isbn={9780226580814},
  lccn={94002418},
  series={Women in Culture and Society Series},
  year={1994},
  publisher={University of Chicago Press},
}

@article{Yan2020,
title = {Buckling of pressurized spherical shells containing a through-thickness defect},
journal = {Journal of the Mechanics and Physics of Solids},
volume = {138},
pages = {103923},
year = {2020},
issn = {0022-5096},
doi = {10.1016/j.jmps.2020.103923},
author = {Dong Yan and Matteo Pezzulla and Pedro M. Reis}
}

@article{Haughton1979,
title = {Bifurcation of inflated circular cylinders of elastic material under axial loading—I. Membrane theory for thin-walled tubes},
journal = {Journal of the Mechanics and Physics of Solids},
volume = {27},
number = {3},
pages = {179-212},
year = {1979},
doi = {10.1016/0022-5096(79)90001-2},
author = {D.M. Haughton and R.W. Ogden},
}

@book{Ogden1985,
author = {Ogden, R. W.},
title = {Non-Linear Elastic Deformations},
publisher={Courier Corporation},
isbn={9780486696485},
year = {1985}
}

@article{Haughton1979_2,
author = {Haughton, D.M. and Ogden, R.W.},
journal = {Journal of the Mechanics and Physics of Solids},
number = {5},
pages = {489-512},
title = {Bifurcation of inflated circular cylinders of elastic material under axial loading—II. Exact theory for thick-walled tubes},
volume = {27},
year = {1979},
doi = {10.1016/0022-5096(79)90027-9},
}

@book{TimoshenkoGere1961,
author = {Timoshenko, S.P. and Gere, J.M.},
publisher = {McGraw-Hill},
title = {Theory of elastic stability. 2.ed.},
year = {1961},
}

@book{Singer1998,
author = {Singer, J and Arbocz, Johann and Weller, Tom},
isbn = {0-470-17298-3},
publisher = {John Wiley \& Sons Incorporated},
title = {Buckling Experiments, Basic Concepts, Columns, Beams and Plates: Vol. 1},
year = {1998},
}

@article{Ding2014,
number = {2},
pages = {378-387},
publisher = {William Wesley and Son},
title = {Pressure–volume curves: revisiting the impact of negative turgor during cell collapse by literature review and simulations of cell micromechanics},
volume = {203},
year = {2014},
author = {Ding, Yiting and Zhang, Yanxiang and Zheng, Quan‐Shui and Tyree, Melvin T},
journal = {The New phytologist},
doi = {10.1111/nph.12829},
}

@article{Borsuk2024,
    author = {Borsuk, Aleca M and Randall, Josh M and Richburg, Jennifer and Montes, Kyra G and Edwards, Erika J and Brodersen, Craig R},
    title = {Palisade cell geometry in relation to leaf optical and photosynthetic properties in Viburnum},
    journal = {Plant Physiology},
    volume = {198},
    number = {4},
    year = {2024},
    doi = {10.1093/plphys/kiae659},
}

@article{Bidhendi2018,
author = {Bidhendi, Amir J. and Geitmann, Anja},
journal = {Plant physiology (Bethesda)},
number = {1},
pages = {41-56},
publisher = {American Society of Plant Biologists},
title = {Finite Element Modeling of Shape Changes in Plant Cells},
volume = {176},
year = {2018},
doi={10.1104/pp.17.01684}
}

@article{Sanati2013,
author = {Nezhad, Amir Sanati and Naghavi, Mahsa and Packirisamy, Muthukumaran and Bhat, Rama and Geitmann, Anja},
journal = {Lab on a chip},
number = {13},
pages = {2599-268},
publisher = {Royal Soc Chemistry},
title = {Quantification of the Young's modulus of the primary plant cell wall using Bending-Lab-On-Chip (BLOC)},
volume = {13},
year = {2013},
doi={10.1039/c3lc00012e}
}

@article{zhang2024biomechanics,
  title={The biomechanics of turgor pressure},
  author={Zhang, Xinyi and Ramakanth, Karthikbabu Kannivadi and Long, Yuchen},
  journal={Current Biology},
  volume={34},
  number={20},
  pages={R986--R991},
  year={2024},
  publisher={Elsevier}
}

@article{pittermann2006mechanical,
  title={Mechanical reinforcement of tracheids compromises the hydraulic efficiency of conifer xylem},
  author={Pittermann, Jarmila and Sperry, John S and Wheeler, James K and Hacke, Uwe G and Sikkema, Elzard H},
  journal={Plant, Cell \& Environment},
  volume={29},
  number={8},
  pages={1618--1628},
  year={2006},
  publisher={Wiley Online Library}
}

@article{Tsegaw2005,
author = {Tsegaw, T and Hammes, S and Robbertse, J},
journal = {HortScience},
number = {5},
pages = {1343-1346},
publisher = {American Society for Horticultural Science},
title = {Paclobutrazol-induced leaf, stem, and root anatomical modifications in potato},
volume = {40},
year = {2005},
doi={https://doi.org/10.21273/HORTSCI.40.5.1343},
}

@article{Sancho2011,
author = {Sancho Knapik, Domingo and Peguero Pina, José Javier and Gómez Álvarez Arenas, Tomás E and Fernández, V and Gil Pelegrín, Eustaquio},
journal = {Journal of experimental botany},
title = {Relationship between ultrasonic properties and structural changes in the mesophyll during leaf dehydration},
year = {2011},
doi = {https://doi.org/10.1093/jxb/err065},
}

@article{Hoshino2019,
author = {Hoshino, Rina and Yoshida, Yuki and Tsukaya, Hirokazu},
journal = {The Plant journal : for cell and molecular biology},
number = {4},
pages = {738-753},
title = {Multiple steps of leaf thickening during sun‐leaf formation in Arabidopsis},
volume = {100},
year = {2019},
doi={https://doi.org/10.1111/tpj.14467},
}

@article{Munekage2015,
author = {Munekage, Yuri Nakajima and Inoue, Shio and Yoneda, Yuki and Yokota, Akiho},
journal = {Plant, cell and environment},
number = {6},
pages = {1116-1126},
title = {Distinct palisade tissue development processes promoted by leaf autonomous signalling and long‐distance signalling in Arabidopsis thaliana},
volume = {38},
year = {2015},
doi={https://doi.org/10.1111/pce.12466},
}

@article{Egesa2024,
author = {Egesa, Andrew Ogolla and Vallejos, C. Eduardo and Begcy, Kevin},
address = {Switzerland},
journal = {Frontiers in plant science},
pages = {1422814-},
title = {Cell size differences affect photosynthetic capacity in a Mesoamerican and an Andean genotype of Phaseolus vulgaris L},
volume = {15},
year = {2024},
doi ={https://doi.org/10.3389/fpls.2024.1422814},
}

@article{Sasaki2019,
author = {Sasaki, Keisuke and Ida, Yuuki and Kitajima, Sakihito and Kawazu, Tetsu and Hibino, Takashi and Hanba, Yuko T.},
journal = {Scientific reports},
number = {1},
pages = {14121-12},
title = {Overexpressing the HD-Zip class II transcription factor EcHB1 from Eucalyptus camaldulensis increased the leaf photosynthesis and drought tolerance of Eucalyptus},
volume = {9},
year = {2019},
doi={https://doi.org/10.1038/s41598-019-50610-5},
}

@article{Wyka2007,
author = {Wyka, T and Robakowski, P and Zytkowiak, R},
journal = {Tree physiology},
number = {9},
pages = {1293-1306},
title = {Acclimation of leaves to contrasting irradiance in juvenile trees differing in shade tolerance},
volume = {27},
year = {2007},
doi={https://doi.org/10.1093/treephys/27.9.1293},
}

@article{Ivancich2012,
author = {Lencinas, Mar{\'i}a Vanessa and Hern{\'a}ndez, Luis and Ivancich, Horacio Sim{\'o}n and Soler Esteban, Rosina Matilde and Mart{\'i}nez Pastur, Guillermo Jos{\'e} and Lindstrom, Ivone},
journal = {Tree physiology},
number = {5},
pages = {554-564},
title = {Foliar anatomical and morphological variation in Nothofagus pumilio seedlings under controlled irradiance and soil moisture levels},
volume = {32},
year = {2012},
doi={https://doi.org/10.1093/treephys/tps024},
}

@article{Ding2014Diameters,
author = {Ding, Yiting and Zheng, Quan-Shui and Zhang, Yanxiang and He, Chunxia and Xie, Bo},
journal = {Journal of plant growth regulation},
number = {2},
pages = {150-159},
title = {Observation of Apparently Unchanging Mesophyll Cell Diameters Throughout Leaf Ontogeny in Woody Species},
volume = {33},
year = {2014},
doi={https://doi.org/10.1007/s00344-013-9357-1},
}

@article{Nyman1977,
author = {Nyman, Leslie P. and Dengler, Nancy G.},
title = {Cell enlargement during leaf development in Catharanthus roseus},
journal = {Canadian Journal of Botany},
volume = {56},
number = {6},
pages = {592-605},
year = {1978},
doi = {https://doi.org/10.1139/b78-068},
}


\newpage

\onecolumn

\begin{center}
\vspace{0.2cm}\noindent \textbf{\large Supplementary Material for
Self-Buckling of Pressurized Cylindrical Tubes}
\end{center}

\setcounter{equation}{0}
\setcounter{figure}{0}
\setcounter{table}{0}
\setcounter{page}{1}
\makeatletter
\renewcommand{\theequation}{S\arabic{equation}}
\renewcommand{\thefigure}{S\arabic{figure}}

\begin{center}
\vspace{0.2cm}\noindent {Morten Opstrup Andersen, Nikolaj T\o{}nner Osvald Olsen, Diksha Bhola, Aleca Borsuk, Craig Brodersen, Anja Geitmann, Matteo Pezzulla}
\end{center}

\textbf{This PDF includes:}
\begin{itemize}
    \item Appendix A: Palisade mesophyll dimensions
\end{itemize}

\appendix
\section{Palisade mesophyll dimensions}\label{SM:sec1}

\begin{table*}[ht]
\centering
\begin{tabular}{@{}lllll@{}}
\toprule
Species                      & Cell Length (µm) & Cell Radius (µm) & R/L   & Source                                                                                    \\ \midrule
Solanum tuberosum            & 87.60            & 7.45             & 0.085 & \cite{Tsegaw2005}                                     \\
Quercus muehlenbergii        & 142.00           & 8.50             & 0.060 & \cite{Sancho2011}                             \\
Arabidopsis thaliana         & 75.00            & 25.00            & 0.333 & \cite{Hoshino2019}                                   \\
Arabidopsis thaliana         & 73.00            & 22.00            & 0.301 & \cite{Munekage2015}                                \\
Phaseolus vulgaris v. Calima & 57.00            & 9.00             & 0.158 & \cite{Egesa2024}                          \\
Phaseolus vulgaris v. Jamapa & 63.00            & 6.25             & 0.099 & \cite{Egesa2024}                                    \\
Eucalyptus camaldulensis     & 64.30            & 4.85             & 0.075 & \cite{Sasaki2019}                        \\
Abies alba                   & 79.90            & 14.15            & 0.177 & \cite{Wyka2007}                  \\
Picea abies                  & 67.50            & 29.80            & 0.441 & \cite{Wyka2007}                  \\
Fagus sylvatica              & 33.80            & 11.75            & 0.348 & \cite{Wyka2007}                  \\
Acer pseudoplatanus          & 69.20            & 14.05            & 0.203 & \cite{Wyka2007}                  \\
Nothofagus pumilio           & 66.40            & 6.60             & 0.099 & \cite{Ivancich2012}                             \\
Syringa oblata               & 40.00            & 3.50             & 0.088 & \cite{Ding2014Diameters}                               \\
Lonicera maackii             & 30.00            & 2.50             & 0.083 & \cite{Ding2014Diameters}                               \\
Prunus yedoensis             & 24.00            & 2.25             & 0.094 & \cite{Ding2014Diameters}                               \\
Catharanthus roseus          & 72.00            & 15.00            & 0.208 & \cite{Nyman1977} \\
Viburnum dentatum            & 25.64            & 30.72            & 1.198 & \cite{Borsuk2024}                                \\
Viburnum furcatum            & 14.75            & 7.43             & 0.504 & \cite{Borsuk2024}                  \\
Viburnum jucundum            & 25.73            & 11.97            & 0.465 & \cite{Borsuk2024}                  \\
Viburnum opulus              & 23.38            & 14.42            & 0.617 & \cite{Borsuk2024}                  \\
Viburnum trilobum            & 20.08            & 11.41            & 0.568 & \cite{Borsuk2024}                  \\
Viburnum hartwegii           & 28.78            & 16.09            & 0.559 & \cite{Borsuk2024}                  \\
Viburnum japonicum           & 41.17            & 12.50            & 0.304 & \cite{Borsuk2024}                  \\
Viburnum lautum              & 29.14            & 13.78            & 0.473 & \cite{Borsuk2024}                  \\
Viburnum carlesii            & 43.60            & 7.50             & 0.172 & \cite{Borsuk2024}                  \\
Viburnum lantana             & 66.51            & 7.22             & 0.109 & \cite{Borsuk2024}                  \\
Viburnum tinus               & 103.39           & 9.25             & 0.089 & \cite{Borsuk2024}                  \\
Viburnum utile               & 85.79            & 6.64             & 0.077 & \cite{Borsuk2024}                  \\
Viburnum cinnamomifolium     & 46.46            & 9.50             & 0.204 & \cite{Borsuk2024}                  \\
Viburnum davidii             & 86.94            & 9.04             & 0.104 & \cite{Borsuk2024}                  \\
Viburnum propinquum          & 49.05            & 9.46             & 0.193 & \cite{Borsuk2024}                  \\ \bottomrule
\end{tabular}
\caption{Overview of palisade mesophyll dimensions in different plant species.  }
\label{tab:leaves}
\end{table*}

\end{document}